\documentclass{PoS}

\usepackage{epsfig}

\newcommand{\tr}{{\rm Tr}}
\newcommand{\non}{\nonumber}
\def\lsim{\raise0.3ex\hbox{$<$\kern-0.75em\raise-1.1ex\hbox{$\sim$}}}
\def\gsim{\raise0.3ex\hbox{$>$\kern-0.75em\raise-1.1ex\hbox{$\sim$}}}

\PoS{PoS(LAT2005)194}

\title{Heavy two- and three-quark free energies at finite T }

\ShortTitle{Heavy two- and three-quark free energies at finite T }

\author{\speaker{K.~H\"{u}bner}, O.~Kaczmarek and O.~Vogt\\
        University Bielefeld, Germany\\
        E-mail: \email{huebner@physik.uni-bielefeld.de},
        \email{okacz@physik.uni-bielefeld.de},
        \email{vogt@physik.uni-bielefeld.de}
}

\abstract{
  We investigate the behaviour of heavy $QQ$- and $QQQ$-free energies in
  $SU(3)$ pure gauge theory at finite temperature. After stating some general
  properties of the free energies, we discuss the $QQ$-antitriplet and -sextet
  channels as well as free energies in different $QQQ$-channels. We perform a
  detailed finite volume scaling analysis of $QQ$-free energies below and above
  $T_c$. Moreover, we also discuss the relation of the $QQ$-antitriplet to the
  $Q\bar Q$-singlet free energies. A comparison of the effective running
  coupling indicates that the $r$-dependence of the free energies differs by a factor of two in
  agreement with the different Casimir factors. We observe that the baryonic
  free energies can be decomposed into a sum of two particle interactions.
}

\FullConference{XXIIIrd International Symposium on Lattice Field Theory\\
                 25-30 July 2005\\
                 Trinity College, Dublin, Ireland}

\begin{document}

\section{Introduction}
In this work we discuss the behaviour of heavy (static) diquark and baryonic systems at
finite temperatures in the quenched approximation. 
We have calculated static diquark and baryonic free energies from Polyakov loop correlation
functions. The results of the diquark free energies are compared to free energies of quark anti-quark
pairs and to the baryonic systems. 

\section{$QQ$ Free Energies}

\subsection{Observables and Simulation Details}
\label{qq_observables}
A system of two quarks can either be in a colour antitriplet (anti-symmetric) or colour
sextet (symmetric) state.
For a static diquark system the free energies in the different colour
channels can be calculated  using 
correlation functions of the  Polyakov loop.
For the  antitriplet and sextet state these are given by
%\cite{Nadkarni:1986cz,Nadkarni:1986as}
\cite{Nadkarni:1986cz}
\begin{eqnarray}
  C^{\bar 3}_{qq}(r,T) & = & \frac{3}{2}\langle\tr L(0)\tr L(r)\rangle -
  \frac{1}{2}\langle\tr L(0)L(r)\rangle\label{eq:corr_fct1}\\\non &&\\
\label{eq:corr_fct2}
  C^6_{qq}(r,T) & = & \frac{3}{4}\langle\tr L(0)\tr L(r)\rangle +
  \frac{1}{4}\langle\tr L(0)L(r)\rangle.
\end{eqnarray}
and the corresponding static diquark free energies are obtained by  
\begin{equation}
F^x_{qq}(r,T)=-T\ln
C^x_{qq}(r,T)\quad\mbox{with}\quad x \in \{{\bar 3},6\}.
\label{eq:fx}
\end{equation}
The free energies calculated in this way describe the change in free energy of
a gluonic system containing a diquark pair in a specific colour channel.
Both correlation functions are gauge dependent quantities. Following
\cite{Philipsen:2002az} we fix to Coulomb gauge. Furthermore they are not Z(3)
symmetric and therefore vanish when averaging over the Z(3) sectors. 
We rotate the configurations such that the Polyakov loop
(averaged over one configuration) lies in the real Z(3) sector. 
We have performed quenched calculations on lattices with various spatial extent
using the tree level Symanzik-improved gauge
action. 
We compare our results with the heavy quark-antiquark free energies calculated
in \cite{Kaczmarek:2002mc,Kaczmarek:2004gv}.

\subsection{General Properties}
\begin{figure}[t]
  \epsfig{file=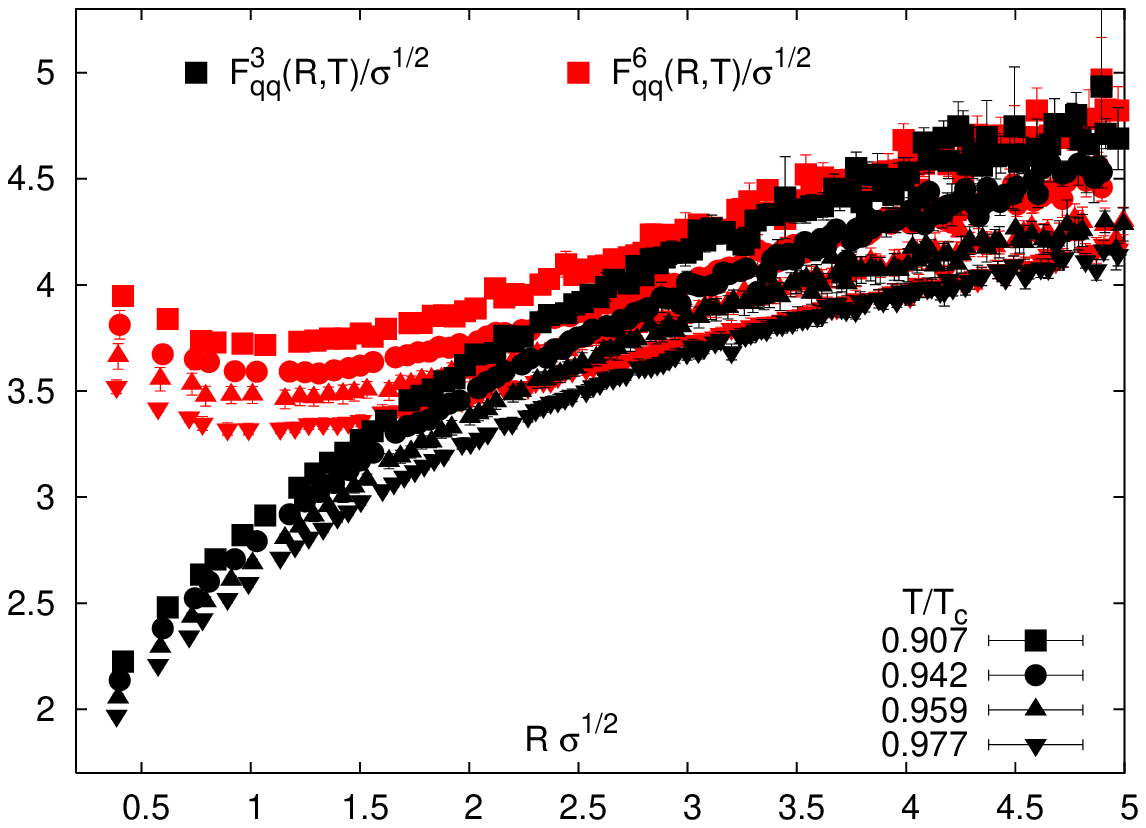,width=7.5cm}
  \epsfig{file=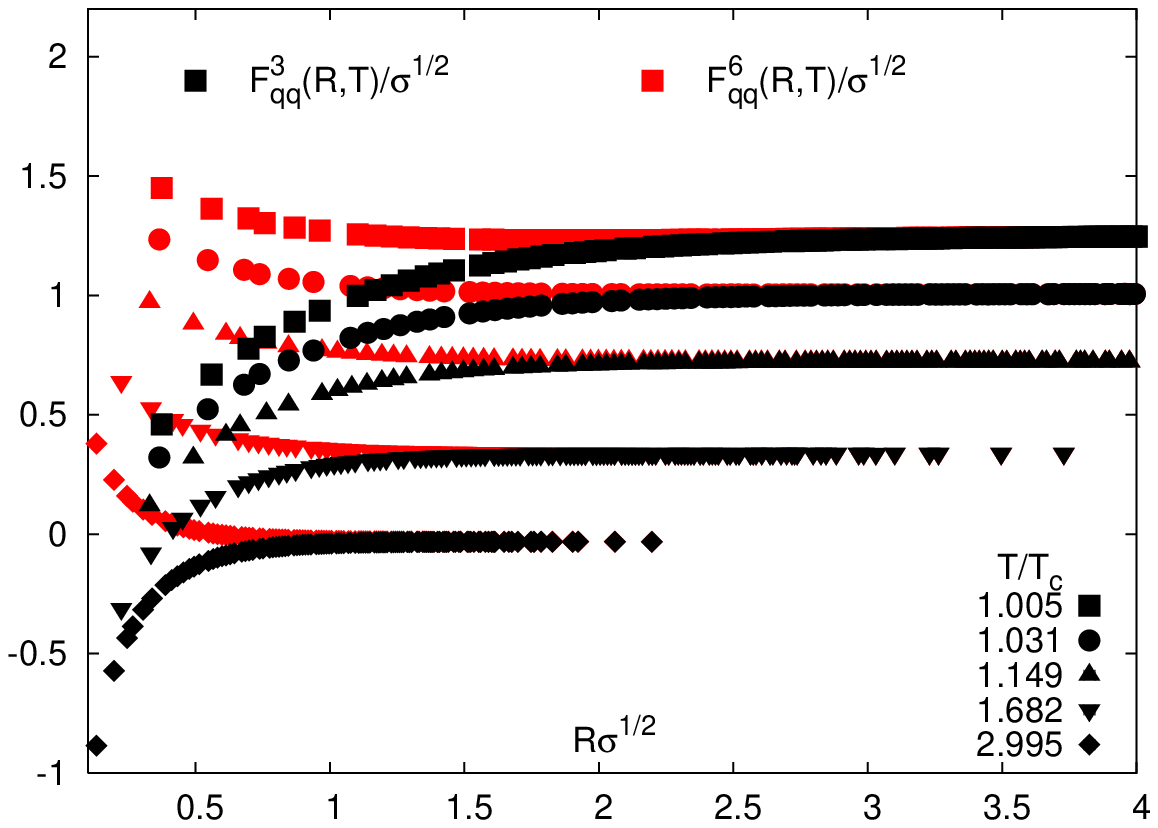,width=7.5cm}
  \caption{\label{fig:plcqq36_3204} Free energies of a static diquark system in
  the antitriplet (black) and sextet (red) channel
  below (left) and above (right) $T_c$ on a
  $32^3\times 4$ lattice.
  }
\end{figure}
In the following the static diquark free energies 
are renormalized
according to the method described in \cite{Kaczmarek:2002mc} by renormalizing
the Polyakov loops using the same
renormalization constants as calculated from the quark-antiquark free energies.

Figure \ref{fig:plcqq36_3204}(left) shows $F^{\bar 3}_{qq}(r,T)/\sqrt{\sigma}$ and
$F^6_{qq}(r,T)/\sqrt{\sigma}$
for four different temperatures below $T_c$. 
We observe clearly $F^3_{qq}(r,T)/\sqrt{\sigma}$ to be attractive over the entire distance
interval calculated here, 
reaching an almost linear behaviour for large distances.
Unlike $Q\bar Q$-singlet free energies at finite $T$,
$F^{\bar 3}_{qq}(r,T)/\sqrt{\sigma}$ retains a $T$-dependence for small distances.    
$F^6_{qq}(r,T)/\sqrt{\sigma}$ is attractive for
distances $r\sqrt{\sigma}\gsim 1.0$, at $r\sqrt{\sigma}\approx  1.0$ it has a minimum and becomes
repulsive for even smaller distances, remaining  $T$-dependent as well.
This qualitative behaviour is already known from $Q\bar Q$-octet free energies.  
For large distances $F^{\bar 3}_{qq}(r,T)/\sqrt{\sigma}$ and $F^6_{qq}(r,T)/\sqrt{\sigma}$ coincide.

In fig.~\ref{fig:plcqq36_3204}(right) we show the antitriplet, $F^{\bar 3}_{qq}(r,T)$, and
the sextet free energies, $F^6_{qq}(r,T)$, for five temperatures above $T_c$. 
While $F_{qq}^{\bar 3}(r,T)$ is attractive over the entire distance range, the $QQ$-sextet free energies are repulsive for all distances. At large separations the
free energies in both colour channels tend towards the same asymptotic value. In
this limit the colour sources get screened independently of their colour
orientation.

\subsection{Thermodynamic Limit}
\begin{figure}[t]
  \epsfig{file=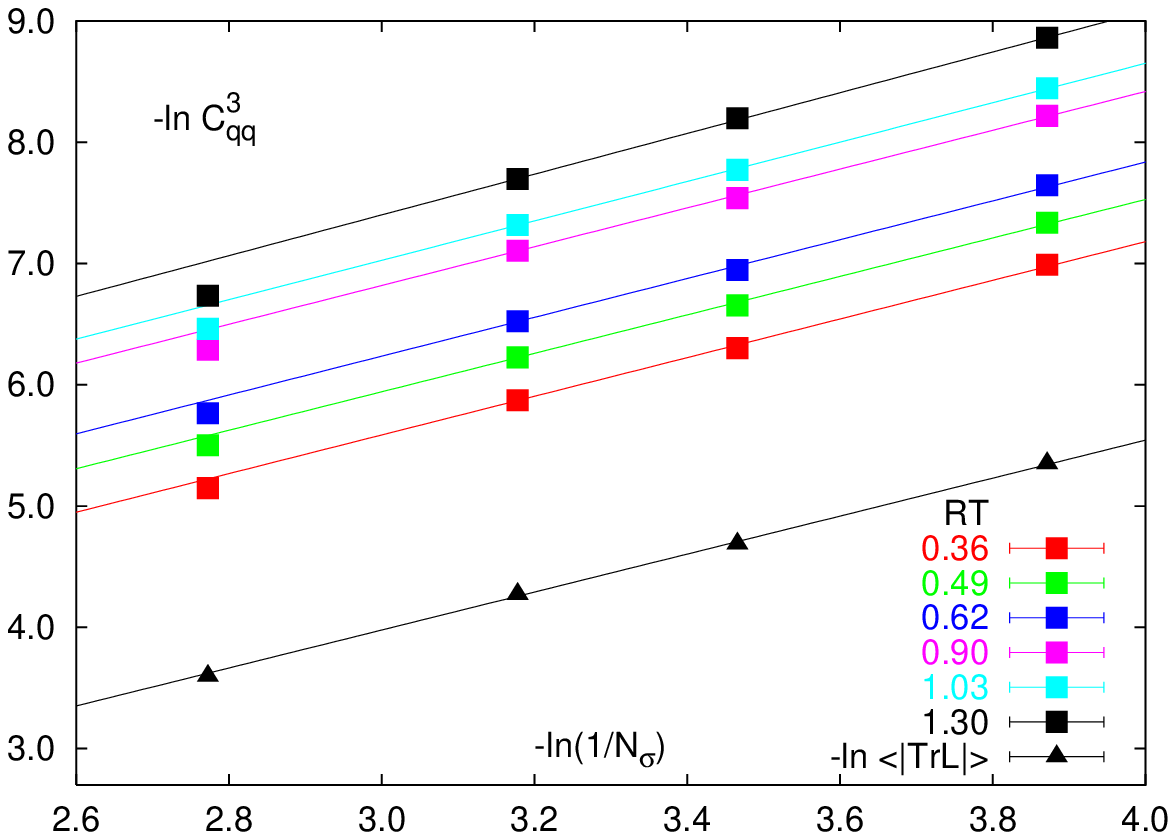,width=7.5cm}
  \epsfig{file=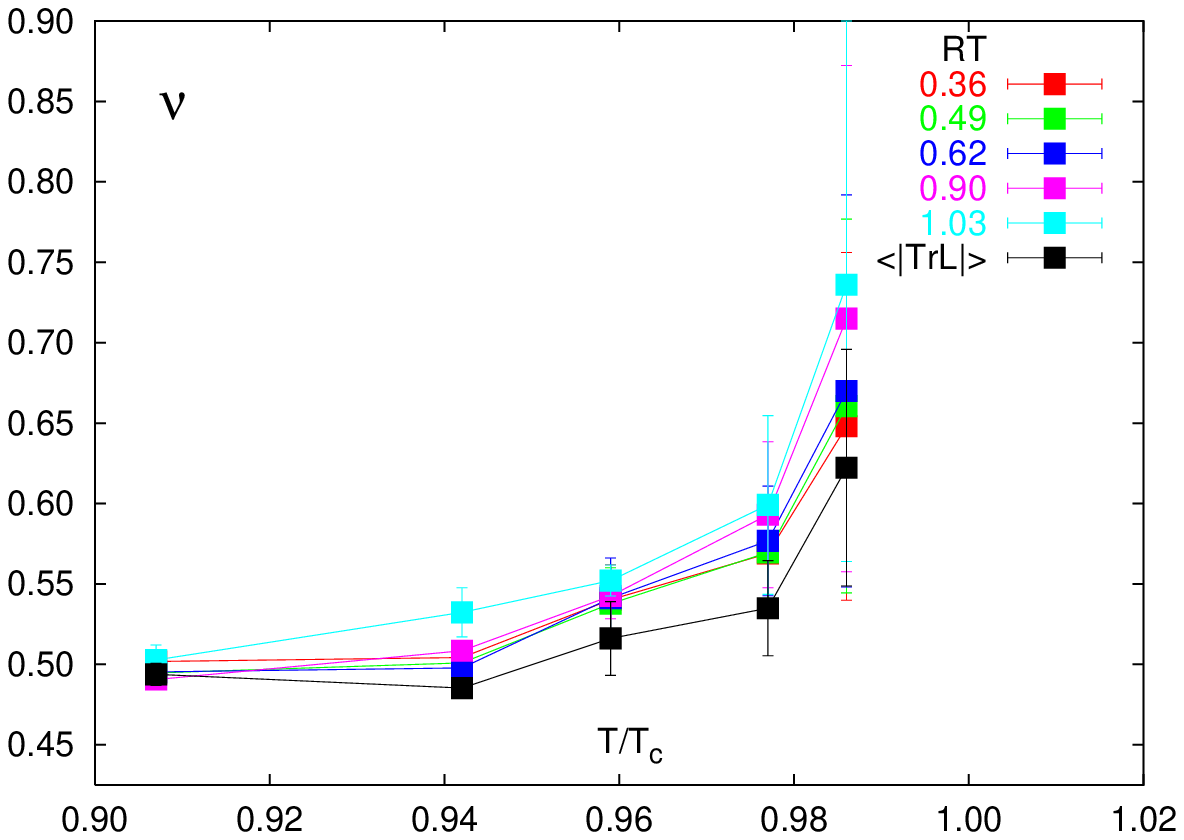,width=7.5cm}
   \caption{\label{fig:plc_scale_u} left: scaling of $C^{\bar 3}_{qq}(r,T)$
  for several distances $rT$ and $\langle|\tr L|\rangle$ with respect to $N_{\sigma}$ for
  $T/T_c=0.959$, right:
  fit parameter $\nu(r,T)$ over $T/T_c$ for  $C^{\bar 3}_{qq}(r,T)$ and $\langle|\tr L|\rangle$.
  }
\end{figure}

As mentioned in \ref{qq_observables}, we rotate the configurations such that the Polyakov loop
(averaged over one configuration) lies in the real Z(3) sector, which for the
Polyakov loop is equivalent to calculating $\langle \vert \tr L \vert \rangle$.
While this expectation value is finite on finite volumes, it vanishes in the
thermodynamic limit. 
In the strong coupling limit its volume dependence can be understood in terms of a random walk 
\cite{engels} that predicts $\langle \vert \tr L \vert \rangle \sim
{V}^{-1/2}$, which we assume to be still valid in the confined phase of
quenched QCD, and furthermore should hold for $C^{\bar 3}_{qq}(r,T)$ as well. 
To check the validity of the scaling behaviour of the Polyakov loop and its
correlation functions we have performed calculations on lattices of size
$N_\sigma^3\times N_\tau$ with fixed temporal extent $N_\tau=4$ and various
spatial extents $N_\sigma=16, 24, 32$ and 48.
The results for $\langle \vert \tr L \vert \rangle$ and $C_{qq}^{3}(r,T)$ at a
temperature of $T/T_c=0.959$ for some
values of the separation $r$ are shown in fig.~\ref{fig:plc_scale_u}(left) and
similar for $T<T_c$. They
clearly show that the expected scaling behaviour holds for all temperatures
below $T_c$ for the Polyakov loop as well as for the correlation functions at
all distances. In the thermodynamic limit all non Z(3) symmetric observables
vanish, despite the fact that we have rotated the gauge field configurations to
a specific Z(3) sector before calculating this observables.
Deviations from a linear behaviour in fig.~\ref{fig:plc_scale_u}(left) are visible for
large distances $r$ on the smallest lattice ($N_\sigma=16$) which may show
the influence of the periodic boundary conditions.\\
We have used the data for a fit of the scaling ansatz
\begin{equation}
 C^{\bar 3}_{qq}(r,T) \propto N_{\sigma}^{-3\nu(r,T)} \propto V^{-\nu(r,T)}.
\end{equation}
The results for the exponent $\nu$ at temperatures below $0.98~T_c$
shown in fig.~\ref{fig:plc_scale_u}(right) are in
agreement with the expected value of 0.5 within errors. The increase near $T_c$
may be explained by an increase of the correlation length near the phase
transition and indicates that the lattice volumes used may be too small near
the critical temperature to observe the correct asymptotic scaling. 
In the deconfined Phase Z(3) symmetry is spontaneously broken and the
expectation values of the Polyakov loop and the $QQ$-correlation functions are
finite in the infinite volume limit.
This is shown in fig.~\ref{fig:plc_scale_o}(left) where no
volume dependence can be identified for all analyzed observables.

\subsection{Comparison to $Q\bar Q$ Free Energies}
In perturbation theory the free energy of a quark-antiquark (diquark) pair are
in leading order given by 
\begin{equation}
  F^1_{q\bar q}(r,T)\simeq -\frac{4}{3}\frac{\alpha(r)}{r}
  \quad\mbox{and}\quad
  F^{\bar 3}_{qq}(r,T)\simeq -\frac{2}{3}\frac{\alpha(r)}{r}
\end{equation}
for $rT\ll 1$ and in high temperature perturbation theory ($rT\gg 1$)
\begin{equation}
  F^1_{q\bar q}(r,T)\simeq -\frac{4}{3}\frac{\alpha(r,T)}{r}e^{-m(T)r}
  \quad\mbox{and}\quad
  F^{\bar 3}_{qq}(r,T)\simeq -\frac{2}{3}\frac{\alpha(r,T)}{r}e^{-m(T)r}.
\end{equation}

\begin{figure}[ht]
  \epsfig{file=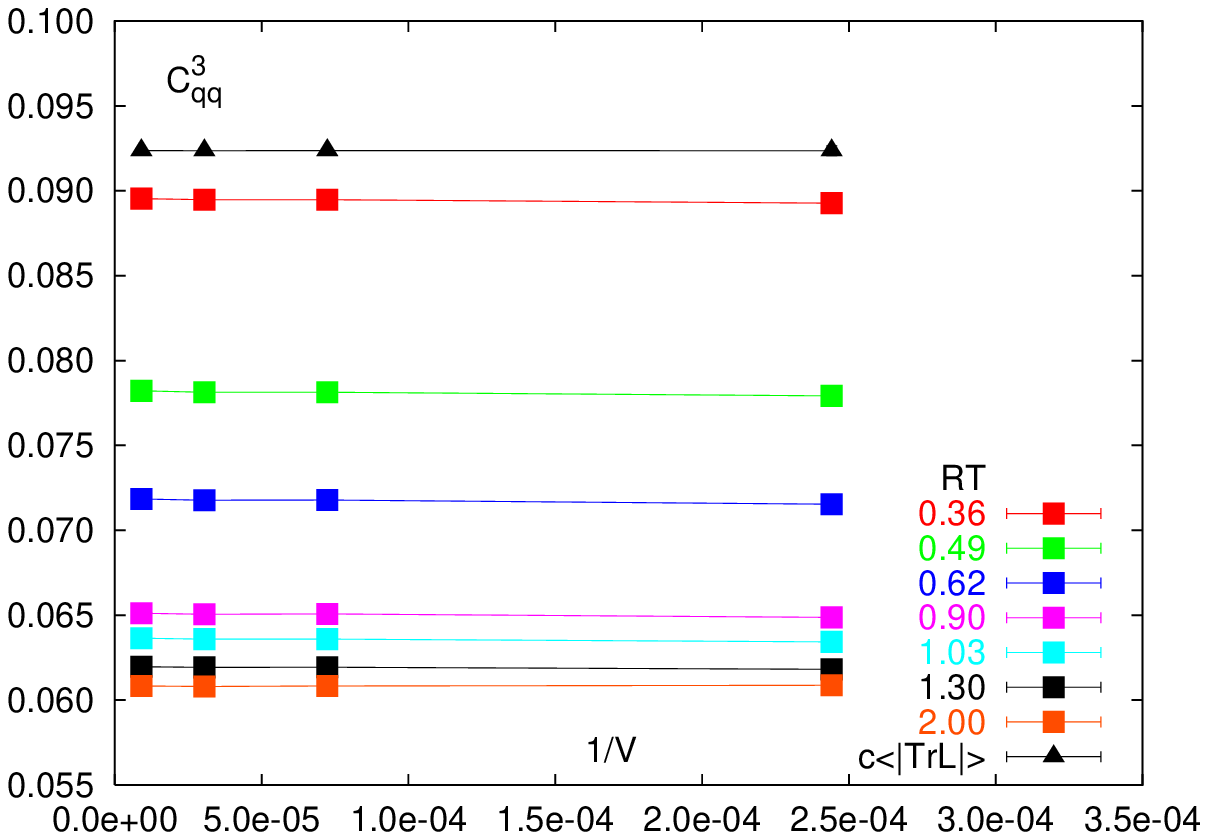,width=7.5cm}
  \epsfig{file=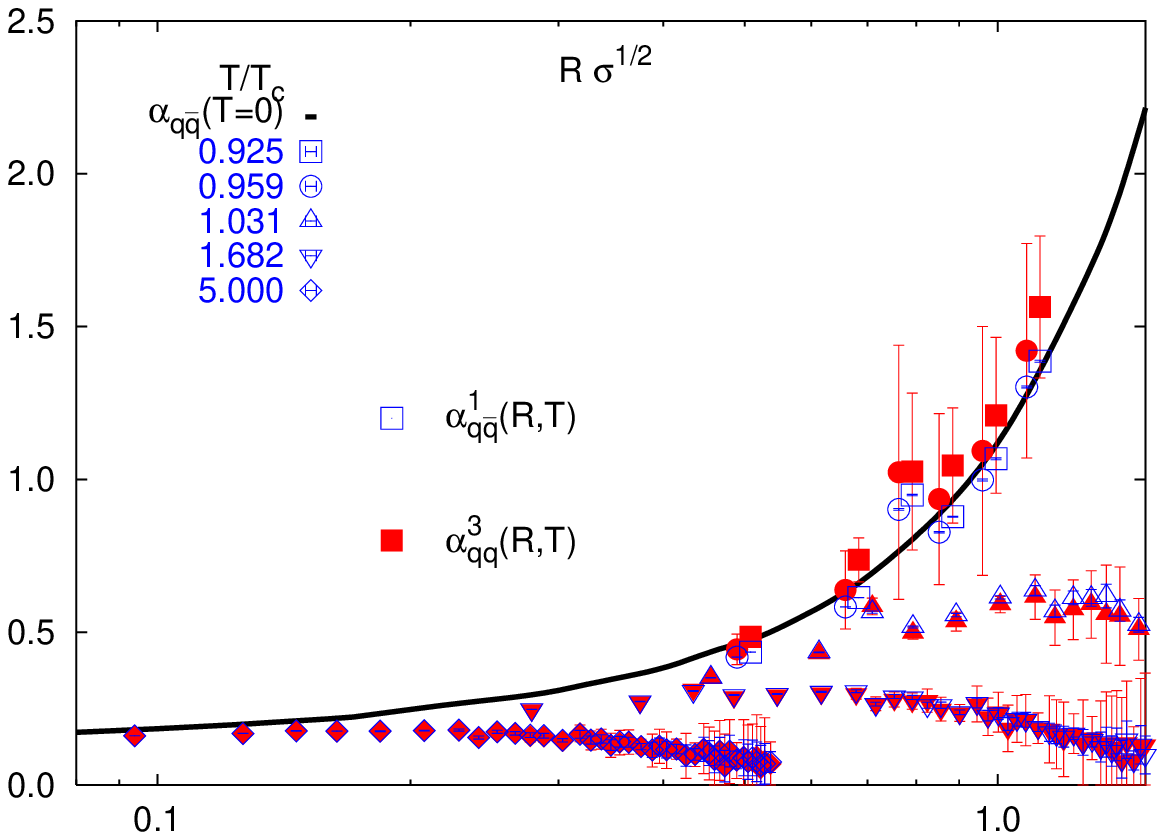,width=7.5cm}
  \caption{\label{fig:plc_scale_o} left: scaling of $C^{\bar 3}_{qq}(r,T)$
  for several distances $rT$ and $\langle|\tr L|\rangle$ (scaled by $c=3/8$) with respect to $1/V$ for
  $T/T_c=1.682$, right: effective couplings $\alpha^1_{q\bar q}(r,T)$ and
  $\alpha^{\bar 3}_{qq}(r,T)$ over $r\sqrt{\sigma}$ on a
  $32^3\times 4$ lattice.
  }
\end{figure}

In order to compare the $Q\bar Q$-singlet and the $QQ$-antitriplet free
energies, we compare the effective couplings of $Q\bar Q$-singlet and the
$QQ$-antitriplet, defined by
\begin{equation}
  \alpha^1_{q\bar q}(r,T)=-\frac{3}{4}r^2\frac{d F^1_{q\bar q}(r,T)}{d r}
  \quad\mbox{and}\quad
  \alpha^{\bar 3}_{qq}(r,T)=-\frac{3}{2}r^2\frac{d F^{\bar 3}_{qq}(r,T)}{d r}.
\end{equation}
In this way all constant contributions are removed.
In  fig.~\ref{fig:plc_scale_o}(right) we show $\alpha^1_{q\bar q}(r,T)$ and $\alpha^{\bar
  3}_{qq}(r,T)$ for temperatures below and above  $T_c$. 
We see, that $\alpha^1_{q\bar q}(r,T)$ and $\alpha^{\bar 3}_{qq}(r,T)$ agree
  within errors up to
  large distances. Therefore we conclude, that $2 \Delta F_{qq}^3(r,T) \simeq \Delta
  F_{q\bar q}^1(r,T)$, where $\Delta F^x := F^x(r,T) - F^x(r\to\infty,T)$.

\section{$QQQ$ Free Energies}

\subsection{Observables}
We have also investigated correlation functions composed of three Polyakov loops
representing a baryonic system of static quarks on the same configurations as
for the diquarks. The colour state of a $QQQ$ system can be in 
an anti-symmetric singlet, two symmetrically mixed octet and in a totally
symmetric decuplet colour state. The observables can be found in
\cite{huebner04} and are gauge dependent quantities, so that we again have
fixed to Coulomb gauge. Here we present only calculations where the three
Polyakov loops are situated at the vertices of an equilateral triangle with the
edge length $R$.

\subsection{General Properties}
In fig.~\ref{fig:plc3_3208}(left) we show the $QQQ$ free energies in the
different colour channels and in the average channel at $6T_c$ from a $32^3\times 8$ lattice.  
We observe that all channels show screening behaviour and approach the same
constant for large distances. The singlet free energy is strongly, the octet
weakly and the average even weaker attractive and the decuplet free energy is
strongly repulsive. For small distances the singlet free energy coincides with
$V_{qqq}(T=0) = 1.5V(T=0)$ becoming $T$-independent like the $Q\bar Q$-singlet
free energies.

\subsection{Comparison to $Q\bar Q$ Free Energies}
\begin{figure}[t]
  \epsfig{file=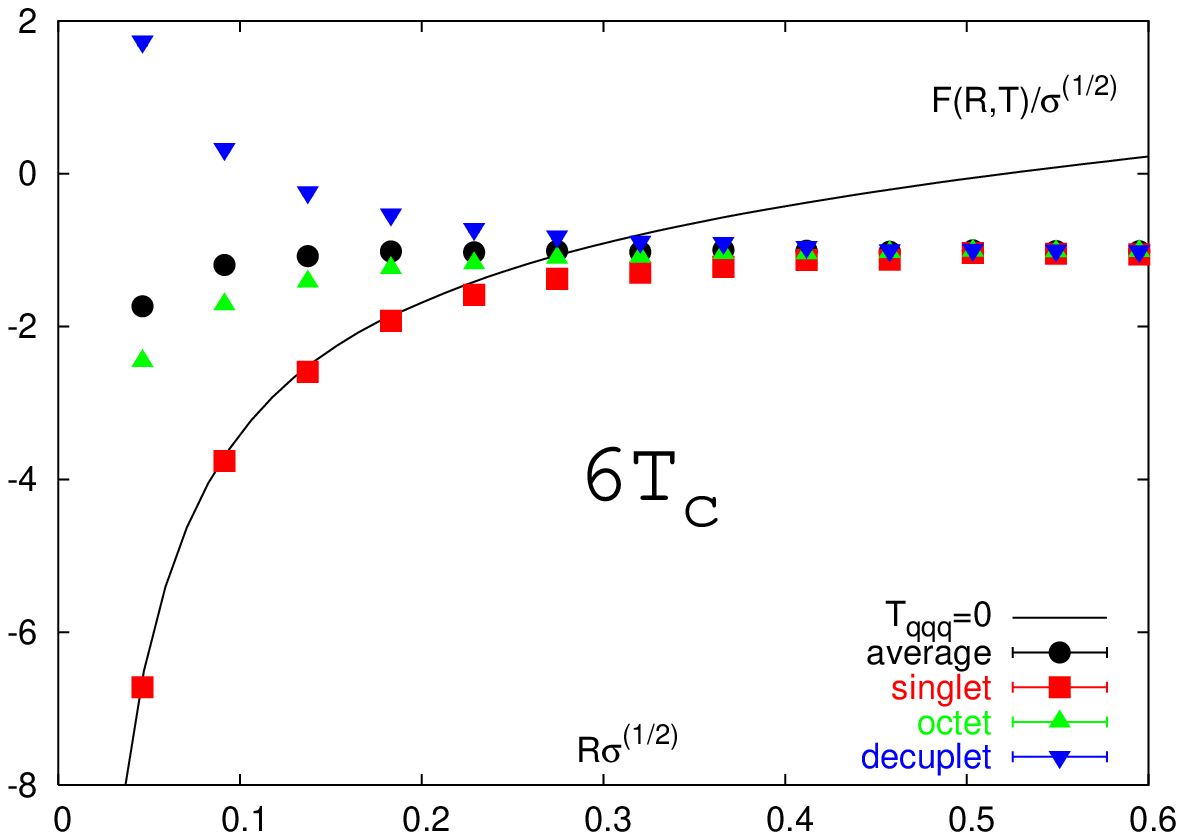,width=7.5cm}
  \epsfig{file=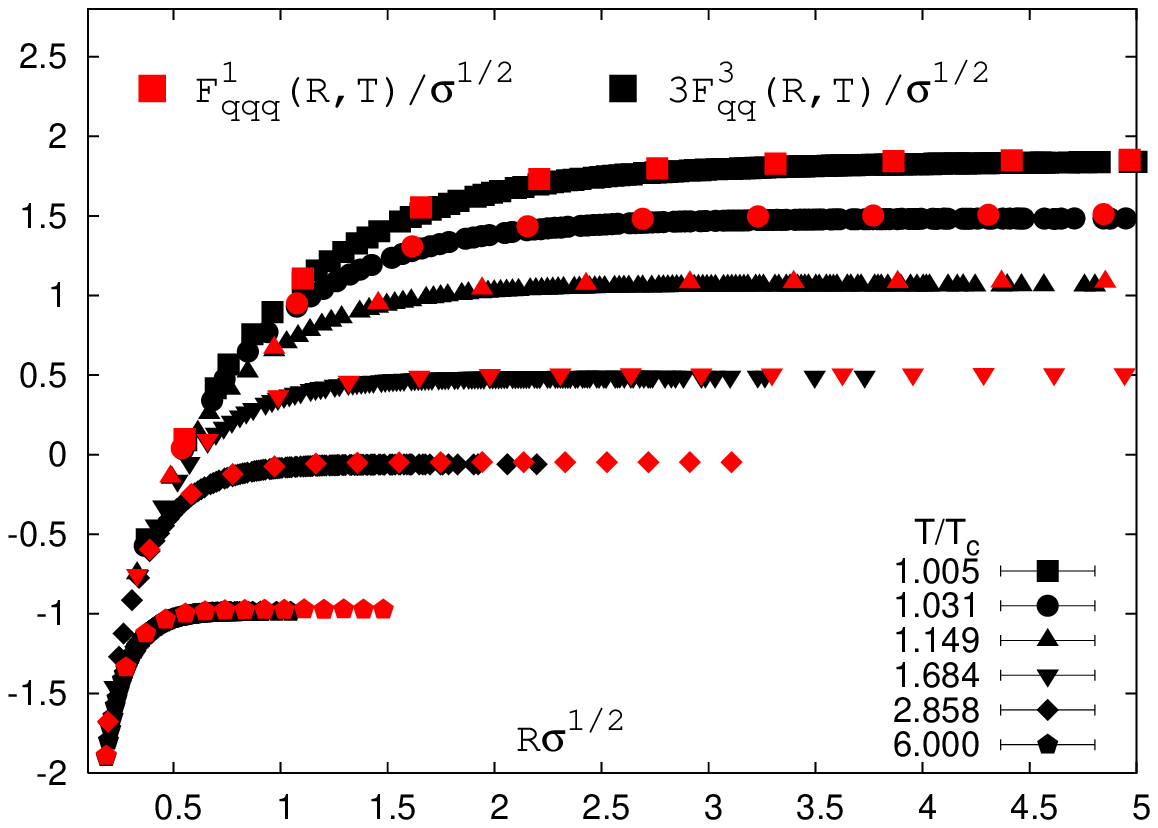,width=7.5cm}
  \caption{\label{fig:plc3_3208} left: Free energies of a static baryonic
  system in the different colour channels at $T/T_c=6$ from a $32^3\times
    8$ lattice, right: comparison of $F^1_{qqq}(r,T)$ (red) and $3F^{\bar
    3}_{qq}(r,T) + 3F_q(T)$ (black) above $T_c$ from a $32^3\times 4$ lattice.
  }
\end{figure}

If one assumes that the interaction of the three static quarks of the baryonic
system is dominated by a sum of two particle interactions, the $QQQ$-singlet
free energy can be expressed by the diquark and quark free energy by  
\begin{equation}
\label{eq:decompos}
  F^1_{qqq}(r,T) = 3F^{\bar 3}_{qq}(r,T) + 3F_q(T).
\end{equation}
In fig.~\ref{fig:plc3_3208}(right) we show the $QQQ$-singlet free energy
$F^1_{qqq}(r,T)$ and three times the $QQ$-antitriplet free energy $F^{\bar
  3}_{qq}(r,T)$ above $T_c$ from
a $32^3\times 4$ lattice compared to the ansatz (\ref{eq:decompos}).
We observe both channels to coincide at all distances and temperatures. %, hence
We therefore conclude that the free energy of the $QQQ$-singlet above $T_c$ as a
function of distance $r$ can be
understood in terms of a sum of two particle interactions of $QQ$ pairs being in the colour
antitriplet state and the solely $T$-dependent free energy of the static  non-interacting
quark $F_q(T)$.      

\section{Conclusion and Outlook}
In this work we discussed the behaviour of heavy (static) diquark and baryonic systems at
finite temperatures in the quenched approximation. 
We have calculated static diquark and baryonic free energies from Polyakov loop correlation
functions in Coulomb gauge. 
We observed the $QQ$-antitriplet and -sextet free energies to show an behaviour 
qualitatively similar to that of $Q\bar Q$-singlet and octet free energies.
We established %the thermodynamic scaling behaviour of 
$C^{\bar 3}_{qq}(r,T)$ and $\langle|\tr L|\rangle$ in the thermodynamic limit to vanish like $1/\sqrt{V}$
below $T_c$ and stay constant above $T_c$. Furthermore we found the effective
couplings  $\alpha^1_{q\bar q}(r,T)$ and $\alpha^{\bar 3}_{qq}(r,T)$ to agree
within errors up to large distances.
We have shown the free energies of the different colour channels of the $QQQ$
system above $T_c$ and compared the $QQQ$-singlet free energy to the
$QQ$-antitriplet free energy and found, that the free energy of the $QQQ$-singlet above $T_c$ as a
function of distance $r$ can be
understood in terms of a sum of two particle interactions of $QQ$ pairs being in the colour
antitriplet state and the solely $T$-dependent free energy of the static  non-interacting
quark.  
 
\section*{Acknowledgements}
This work has been supported by DFG under grant GRK 881-1.

\end{document}